\documentclass[pra,twocolumn,showpacs,floatfix]{revtex4}
\usepackage{graphicx}
\usepackage{color}
\usepackage{amsmath}
\begin{document}

\title{Fragility of the Laughlin state in an anharmonically-trapped Bose-Einstein condensate}

\author{A. Roussou$^1$, J. Smyrnakis$^2$, M. Magiropoulos$^2$, N. K. Efremidis$^1$, W. von Klitzing$^3$, 
and G. M. Kavoulakis$^2$}
\affiliation{$^1$Department of Applied Mathematics, University of Crete, GR-71004,
Heraklion, Greece \\
$^2$Technological Education Institute of Crete, P.O. Box 1939, GR-71004, Heraklion, 
Greece \\
$^3$Institute of Electronic Structure and Laser (IESL), Foundation for Research and 
Technology (FORTH), N. Plastira 100, Vassilika Vouton, 70013, Heraklion, Crete, Greece}

\begin{abstract}

When a Bose-Einstein condensate rotates in a purely harmonic potential with an angular frequency 
which is close to the trap frequency, its many-body state becomes highly correlated, with the most
well-known being the bosonic Laughlin state. To take into account that in a real experiment no 
trapping potential is ever exactly harmonic, we introduce an additional weak, quartic potential 
and demonstrate that the Laughlin state is highly sensitive to this extra potential. Our results 
imply that achieving these states experimentally is essentially impossible, at least for a 
macroscopic atom number. 

\end{abstract}


\maketitle

An interesting problem in the field of cold atomic gases, which has been investigated both theoretically,
see, e.g., Refs.\,\cite{qh1, qh1p, qh2, qh3, qh4, qh4p, qh5, qh6, qh7, qh8, qh9, qh10, qh11, qh12}, as well 
as experimentally \cite{Dalibard, Cornell}, is the realization of highly-correlated states of rotating 
Bose-Einstein condensates trapped in harmonic potentials. Theses states show up in the limit where the 
angular frequency of the rotation of the trap $\Omega$ approaches the trap frequency $\omega$, and the 
centrifugal potential exactly cancels the trapping potential. In this limit there exists even an analytic 
expression for the many-body state, which is the bosonic version of the Laughlin state \cite{LS} that 
appears in the quantum Hall effect \cite{HE, HErev}.  

The theoretical model that leads to these correlated states assumes an exactly harmonic trapping potential. 
However, in a real experiment the trapping potential always exhibits anharmonic corrections. Thus, a question 
that arises naturally is whether these states persist, even for ``weak" deviations from a purely harmonic 
potential. References \cite{qhanh1, qhanh2} have considered such deviations. In particular, motivated by the 
experiment of Ref.\,\cite{Dalibard}, the theoretical study of Ref.\,\cite{qhanh1} has considered the effect 
of an additional, repulsive Gaussian potential, which acts at the trap center. Contrary to the model considered 
in the present study, the potential of Ref.\,\cite{qhanh1} is still harmonic for large distances from the 
center of the trap. In addition, the more mathematically-rigorous studies of Refs.\,\cite{qhanh2} considered 
a quadratic-plus-quartic potential, and values of the angular momentum which go way beyond the one where the 
Laughlin state appears and other correlated states show up. In these studies the authors used trial states 
in order to derive conditions on the parameters of their model for its ground state to be asymptotically 
strongly correlated. Numerous theoretical studies have examined the rotational response of a Bose-Einstein 
condensate that is confined in a quadratic-plus-quartic potential for low rotational frequencies of the trap, 
where the system is well described by the mean-field approximation. They have shown that there are three phases, 
namely a vortex lattice, giant-vortex states, and a ``mixed" phase, i.e., a vortex lattice with giant vortices 
in the middle of the trap, see, e.g., Refs.\,\cite{SF, EL, U, FB, KB, JK, Str, fan, Zar, Corr}. These phases 
appear depending on the value of $\Omega$ and the interaction strength.

In the present study we assume that in addition to a harmonic potential, there is a weak quartic 
potential and focus on its effect on the (bosonic) Laughlin state. We start by evaluating this many-body 
state in a purely harmonic potential, which is our reference state. We then identify the effect of the 
anharmonic part of the potential on the many-body state. We stress that while we have focused on the 
Laughlin state, our results are more general, at least for the states with an angular momentum higher 
than, but of the same order as the Laughlin state. 

In what follows we first introduce our model Hamiltonian, which includes the usual harmonic, and a 
weak quartic trapping potential, while the interatomic interactions are modelled as the usual contact
potential. Since the (bosonic) Laughlin state is highly correlated, we necessarily use the method of 
diagonalization of the many-body Hamiltonian, considering small atom numbers. We focus on the limit of 
rapid rotation and investigate the effect of the quartic potential on the energy, the single-particle 
density distribution, the density matrix, and the pair-correlation function of the evaluated (lowest-energy) 
many-body state. Finally, we calculate the overlap of this many-body state with the Laughlin state and the 
giant-vortex state, and find that there is a competition between them. 

The novelty of our results is thus basically twofold. First of all, our study provides a very clear 
picture of the behaviour of the system in the limit of rapid rotation, in the presence of an anharmonic 
potential and the transition from a correlated state to a mean-field state. Equally important is the 
conclusion that the correlated states, which in purely harmonic potentials are theoretically expected 
for rapidly-rotating Bose-Einstein condensates, are extremely fragile and as a result their experimental 
realization is very difficult.
 
Starting with our model, we consider bosonic atoms which are confined in a plane, via a very tight potential 
in the perpendicular direction, and are also subject to an axially-symmetric trapping potential along their 
plane of motion, $V(\rho)$, where $\rho$ is the radial coordinate. This trapping potential is assumed to be 
anharmonic (we set the atom mass $M$, the trap frequency of the harmonic potential $\omega$, and $\hbar$ equal 
to unity),  
\begin{eqnarray}
 V(\rho) = \frac 1 2 \rho^2 ( 1 + \lambda \rho^2 ),
 \label{trpot}
\end{eqnarray}
with a weak quartic part, i.e., $0 < \lambda \ll 1$ (a negative $\lambda$ would not allow stable trapping 
for $\Omega \to \omega$). Typical lowest values of $\lambda$ are on the order of $\lambda = 0.001$ \cite{Dalibard}. 
The atom-atom interaction is modelled as the usual contact potential, $V_{\rm int} = g_{2D} \delta(\vec{\rho} 
- \vec{\rho'})$, where $g_{2D}$ is the strength of the effective two-body interaction (for the effectively 
two-dimensional problem that we consider). 

When the trapping potential is purely harmonic, $\lambda = 0$, and $\Omega \to 1^-$, the system enters 
the regime of the lowest-Landau-level approximation \cite{qh4}, since it expands radially and the density 
becomes low. The nodeless eigenstates of the harmonic potential are 
\begin{equation}
 \psi_m = \frac 1 {\sqrt{\pi m!}} z^m e^{-|z|^2/2},
 \label{LLL}
\end{equation} 
where $z = \rho \exp(i \phi)$, with $\phi$ being the azimuthal angle in cylindrical polar coordinates and
$m \ge 0$ is the quantum number corresponding to the angular momentum (negative values of $m$ correspond 
to states outside the lowest-Landau-level). Actually, in this limit of rapid rotation the exact eigenstate 
of the many-body Hamiltonian for $L = N (N-1)$, where $L$ is the total angular momentum, is the non-mean-field, 
Laughlin-like state, 
\begin{eqnarray}
  \Psi_L \propto \prod_{i < j =1}^N (z_i - z_j)^2 \exp \left( \sum_{i=1}^N -|z_i|^2/2 \right),
  \label{Laughlin}
\end{eqnarray}
which is built from the states $\psi_m$. The main characteristic of the above state is that it has nodes 
when $z_i = z_j$ and as a result its interaction energy vanishes (i.e., it reaches its lowest bound). 

In our analysis below we still consider weak interatomic interactions, i.e., we assume that the interaction 
energy is much smaller than the oscillator quantum of energy. The additional assumption of a weak anharmonic 
potential ($0 < \lambda \ll 1$) allows us to restrict ourselves to the lowest-Landau-level eigenstates 
of Eq.\,(\ref{LLL}). We stress that the eigenstates of the anharmonic potential are still $\psi_m$ to first 
order in $\lambda$ due to the axial symmetry of the quartic potential, as first-order perturbation theory 
implies. 

Our Hamiltonian in second-quantized form is 
\begin{eqnarray}
 H = \sum_m \epsilon_m a_m^{\dagger} a_m + \frac g 2 \sum_{m,n,k,l} I_{mnkl} \, 
 a_m^{\dagger} a_n^{\dagger} a_k a_l \, 
 \delta_{m+n, k+l},
\end{eqnarray}
where $\epsilon_m$ is the single-particle energy, $g = g_{2D} \int |\psi_0|^4 \, d^2 \rho = g_{2D}/(2 \pi)$, 
and to lowest order in $\lambda$, $I_{mnkl} = {(m+n)!}/ [{2^{m+n} \sqrt{m! n! k! l!}}]$. Here perturbation 
theory implies that $\epsilon_m$ is given by 
\begin{eqnarray}
  \epsilon_m = m + \frac {\lambda} 2 (m+1) (m+2),
\end{eqnarray}
where the first term on the right side comes from the harmonic potential and the second from the quartic. 
In this perturbative approach the corrections to first order in $\lambda$ only appear in the single-particle
energies $\epsilon_m$. While in a purely harmonic potential $\epsilon_m$ scales linearly with $m$, for any 
finite, positive $\lambda$, $\epsilon_m$ has a positive curvature. This has serious consequences, since
the well-known degeneracy of the harmonic potential is lifted \cite{KB, JK}.

In our model there are three energy scales (per particle). The first is the quantum of energy associated 
with the harmonic potential. For the values of $L \sim N^2$ that we consider the energy due to the harmonic 
potential is $\sim N$. The second is the energy associated with the quartic part of the trapping potential, 
which is $\sim N^2 \lambda$. The third energy scale is the one associated with the interaction energy, which 
is $\sim (N-1) g$. The restriction to the lowest-Landau level eigenstates requires that $g$ and $\lambda N$
should be at most of order 1.

We now turn to our results. We start with the lowest-energy eigenvalue of the many-body Hamiltonian for some 
given value of $L$ and for the case of a purely harmonic potential ($\lambda = 0$). In Fig.\,1 we consider 
$N = 5$ atoms, with $g = 0.1$ and a truncation $0 \le m \le 8$. From Eq.\,(\ref{Laughlin}) it follows that 
$m_{\rm max}$ has to be at least $2 (N-1) = 8$, for $N = 5$. This plot shows the lowest eigenenergy in 
the rotating frame, $E' = E - L \Omega$, for $\Omega = 1$ from $L=0$, up to $L = 21$. (For $\lambda = 0$ and 
$\Omega = 1$, $E'$ is also the interaction energy.)  

For $\lambda = 0 $ the spectrum has the expected features. For example, it is exactly linear for $2 \le 
L \le N$, see, e.g., \cite{KMP, Thomas, JKlin}, etc. For $L = N (N-1) = 20$ we get the Laughlin state, 
whilst for even higher values of $L$ there are other correlated states. For $L \ge 20$ the interaction 
energy vanishes exactly, in agreement with our numerical results. 

To see the effect of the quartic potential, we also plot in Fig.\,1 the result of the same calculation
for $\lambda = 0.05$. Its effect is drastic, and already for such a small value of $\lambda$ one finds
a distinctly different spectrum. The most distinct feature is that it starts to develop a quasi-periodic 
behaviour, as in a ring potential (in a ring trapping potential the spectrum is periodic on top of a 
parabola, as Bloch's theorem implies \cite{FBl}), developing local minima when $L$ is an integer multiple 
of $N$, i.e., for $L = 5, 10, 15$, and 20. This result, as well as the ones that follow below, show a 
transition of the many-body state from the (correlated) Laughlin state to a (mean-field) giant-vortex 
state as $\lambda$ increases. 

\begin{figure}[h]
\includegraphics[width=9.5cm,height=4.8cm,angle=0]{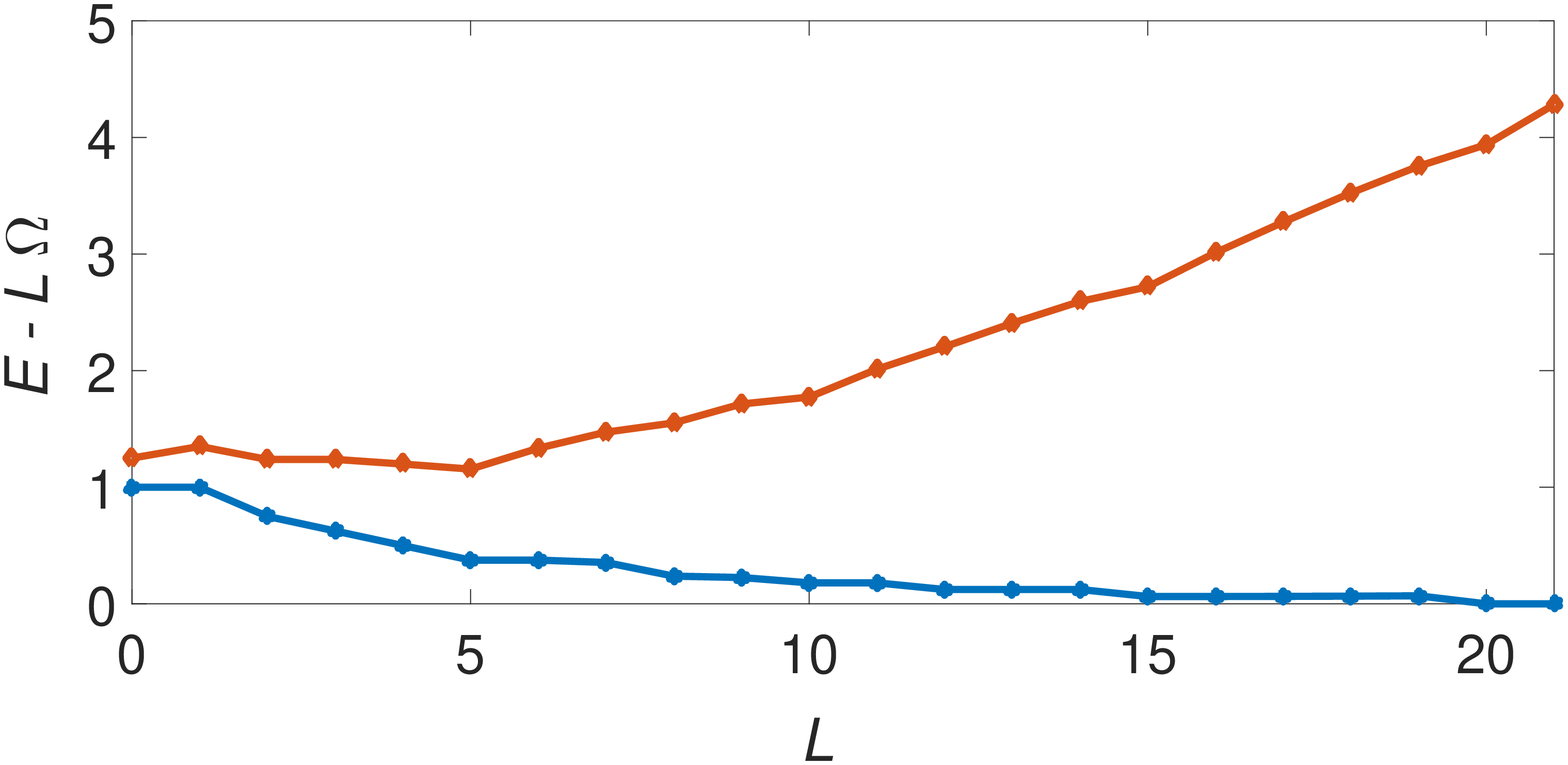}
\caption{(Color online) The lowest-energy eigenvalue of the many-body Hamiltonian in the rotating frame as a
function of $L$, for $\Omega = 1$, $N = 5$ atoms and $g = 0.1$, in a purely harmonic potential, $\lambda = 0$
(lower) and in an anharmonic potential with $\lambda = 0.05$ (higher). For $\lambda = 0$ and $L = N (N-1) = 20$ 
we have the Laughlin state, given in Eq.\,(\ref{Laughlin}).}
\end{figure}

\begin{figure}[h]
\includegraphics[width=9.5cm,height=4.8cm,angle=0]{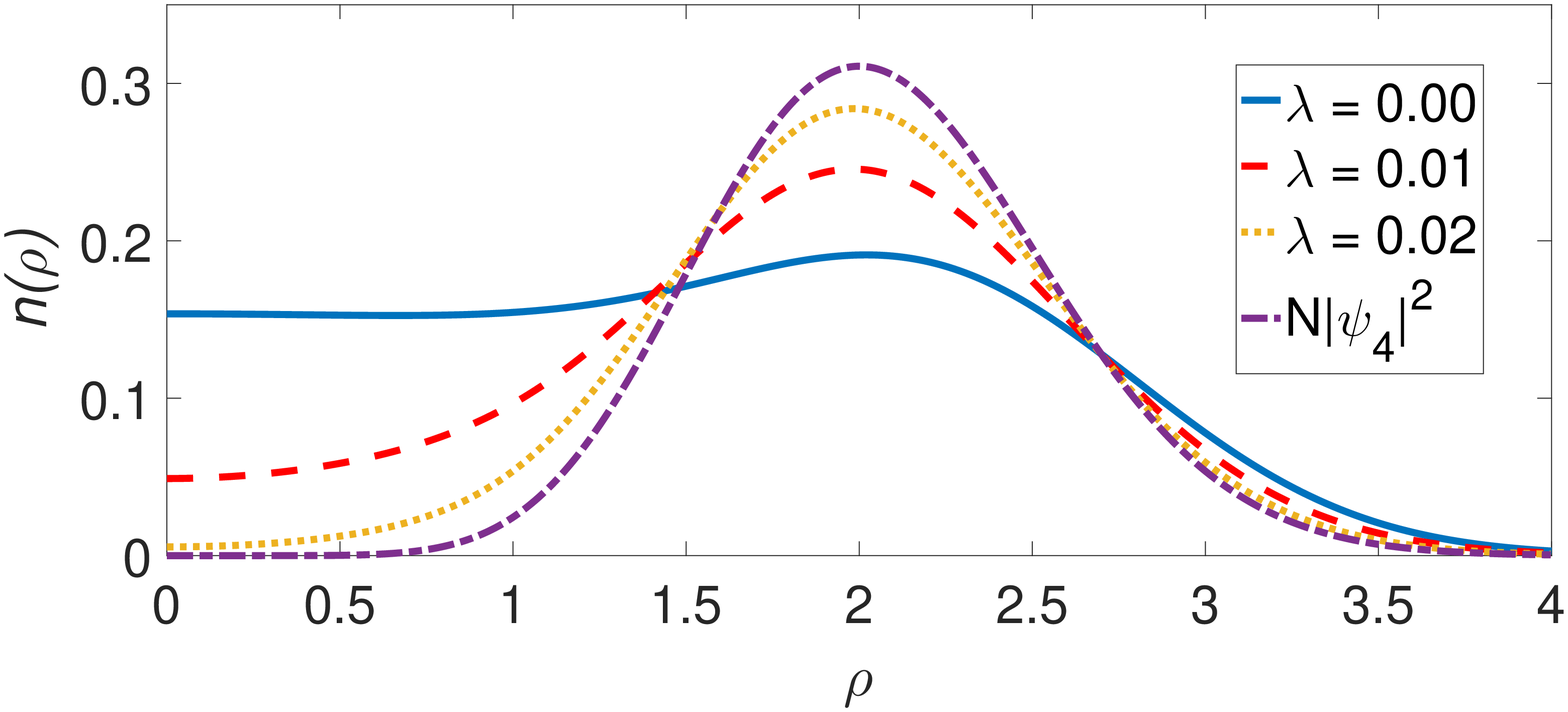}
\caption{(Color online) The single-particle density distribution $n(\rho)$ of the lowest-energy eigenstate 
of the Hamiltonian, for $N = 5$ atoms, $L = 20$, and $g = 0.1$, for $\lambda = 0.00$ (solid line), 0.01 
(dashed line), and 0.02 (dotted line). With increasing $\lambda$, $n(\rho)$ approaches $N |\psi_4(\rho)|^2$, 
which is also shown in the plot (dotted-dashed line).}
\end{figure}

\begin{figure}[h]
\includegraphics[width=9.5cm,height=4.8cm,angle=0]{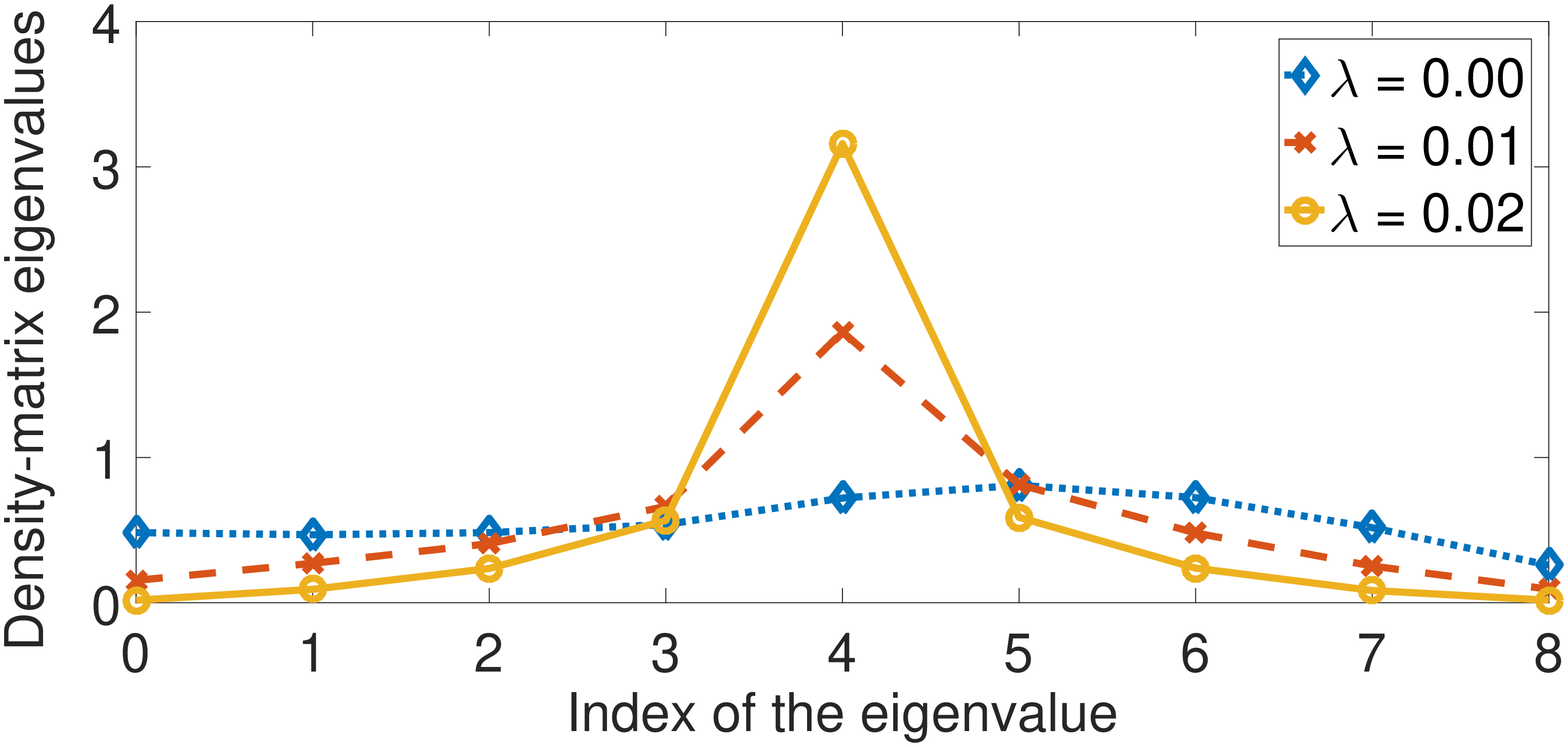}
\caption{(Color online) The eigenvalues of the density matrix of the lowest-energy eigenstate of the 
Hamiltonian, for $N=5$, $L=20$, $g=0.1$, and $\lambda = 0.00$ (dotted line), 0.01 (dashed line), and 0.02 
(solid line).}
\end{figure}

\begin{figure}[h]
\includegraphics[width=9.5cm,height=4.8cm,angle=0]{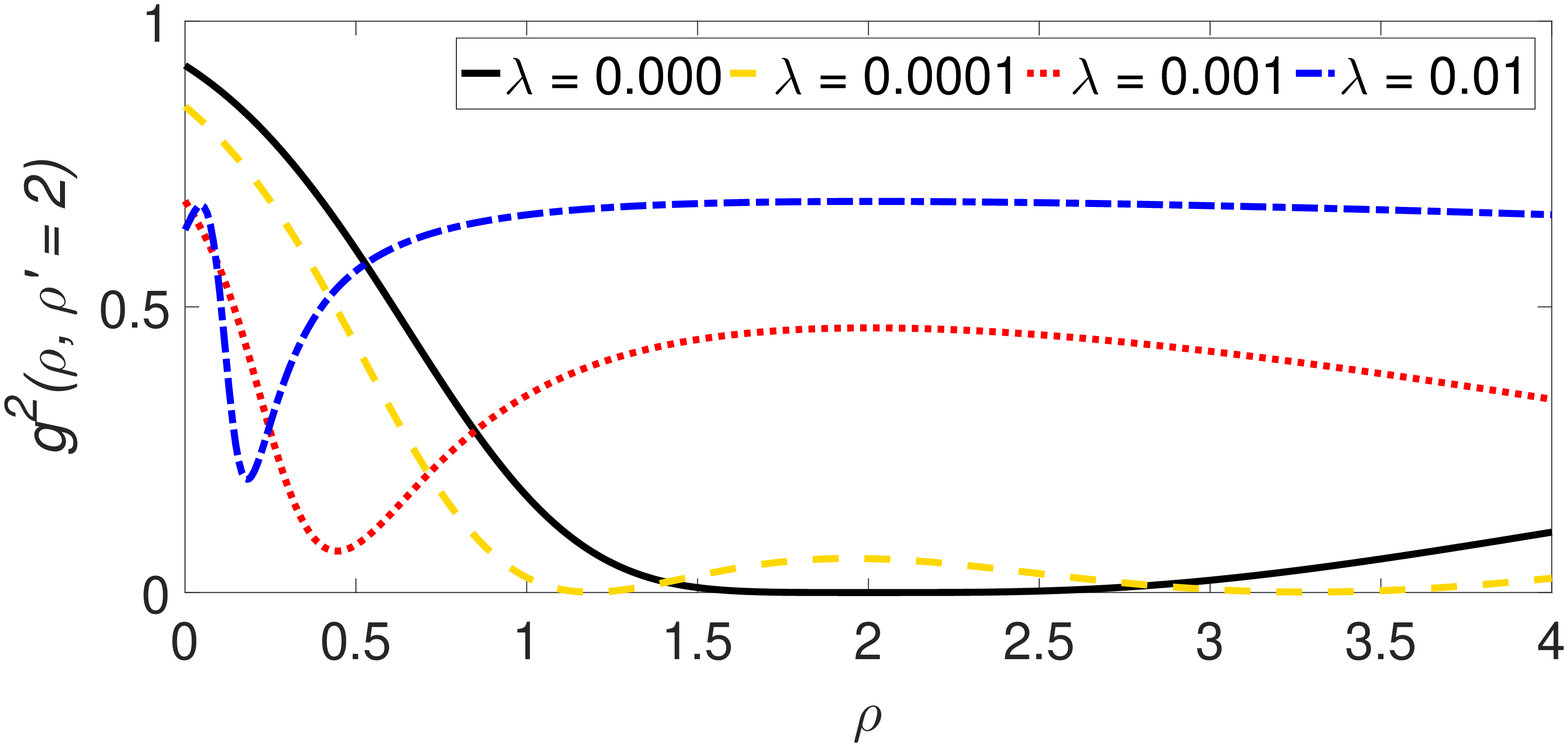}
\caption{(Color online) The pair-correlation function $g^{(2)}(\rho, \rho' = 2)$ for the lowest-energy 
eigenstate of the Hamiltonian, for $N = 4$ atoms, $L = 12$, and $g = 0.001$, for $\lambda = 0$ (solid line), 
0.0001 (dashed line), 0.001 (dotted line), and 0.01 (dotted-dashed line).}
\end{figure}

Let us now turn to the (axially-symmetric) single-particle density, $n(\vec{\rho}) = \langle \Phi^{\dagger}(\vec{\rho}) 
\Phi(\vec{\rho}) \rangle$, where $\Phi(\vec{\rho})$ is the operator that destroys a particle at $\vec{\rho}$. 

Figure 2 shows the single-particle density $n(\rho)$ for $N = 5$ atoms, $L = 20$, $g = 0.1$ and for three 
values of $\lambda = 0.00$, 0.01, and 0.02. For $\lambda = 0.00$, i.e., for the Laughlin state, we see that 
indeed the density is roughly constant and close to the expected result $1/(2 \pi) \approx 0.16$, while the 
radius is also close to the expected result $\sqrt{2 N} = \sqrt{10} \approx 3.16$, to leading order in $N$
\cite{LS}.

It is seen clearly that the effect of the anharmonic potential is to create a ``hole" in the middle of the 
cloud. Actually, even for the rather small value of $\lambda = 0.02$, the single-particle density is well
approximated by that of the ``giant vortex", $N |\psi_4(\rho)|^2$, as seen also in Fig.\,2. The maximum of
this is at $\rho = \sqrt{N-1} = 2$, with a value $N/\sqrt{N-1}/(\sqrt{2 \pi^3}) \approx 0.32$. 

Further evidence of this transition is also seen from the the eigenvalues of the density matrix, $\langle
a_m^{\dagger} a_n \rangle$ (which coincide with the occupancies of the single-particle states, since the 
density matrix is diagonal, due to the axial symmetry of the problem). Figure 3 shows this result for $N=5$, 
$L=20$, $g=0.1$, and for the same values of $\lambda$ considered in Fig.\,2, i.e., 0.00, 0.01, and 0.02. As
$\lambda$ increases we observe that the occupancy of the single-particle state $\psi_{m_0}$, with $m_0 = L/N 
= 4$, becomes dominant, which is consistent with the transition to a giant-vortex state. 

In both plots the transition from the Laughlin state to the giant vortex takes place when the energy (per atom) 
due to the quartic part of the potential, $\lambda N^2$ is comparable with the interaction energy of the giant
vortex. This is $\sim g n_{2D}$, where the two-dimensional density $n_{2D} \sim N/R \sim \sqrt{N}$. Here $R$ 
is the radius of the (roughly) homogeneous density of the Laughlin state, which is $\propto \sqrt{N}$. Thus the 
``threshold" value of $\lambda$ is $\sim g/N^{3/2} \sim 0.01$, in agreement with the results of Figs.\,2 and 3.

Another relevant quantity is the pair-correlation function, which is defined as
\begin{eqnarray}
 g^{(2)}(\vec{\rho}, \vec{\rho'}) = \frac {\langle \Phi^{\dagger}(\vec{\rho}) \Phi^{\dagger}(\vec{\rho'}) 
 \Phi(\vec{\rho'}) \Phi(\vec{\rho}) \rangle} 
 {\langle \Phi^{\dagger}(\vec{\rho}) \Phi(\vec{\rho}) \rangle 
 \langle \Phi^{\dagger}(\vec{\rho'}) \Phi(\vec{\rho'}) \rangle}. 
\end{eqnarray}
In an uncorrelated, mean-field, state $g^{(2)}(\vec{\rho}, \vec{\rho'})$ is a straight line and equal to 
$(N-1)/N$.

Figure 4 shows $g^{(2)}(\rho, \rho' = 2)$, with $\vec{\rho}$ and $\vec{\rho'}$ pointing at the same direction,  
for $N = 4$ atoms, $L = 12$, $g = 0.001$ and $\lambda = 0, 0.0001, 0.001$, and 0.01. The reference point is chosen 
to be at $\rho' = 2$ (this is roughly where the maximum of the single-particle density distribution is located). 
That is why for $\lambda = 0$ there is a node in $g^{(2)}(\rho, \rho' = 2)$ at $\rho = 2$, as expected from the 
Laughlin state. As $\lambda$ increases the node disappears. In addition, for values of $\rho$ larger than 2, 
$g^{(2)}(\rho, \rho' = 2)$ is roughly constant (differing from 3/4 due to the finiteness of $N$ and the 
relatively small values of $\lambda$), while a local minimum forms at some value of $\rho$ which is smaller 
than $\rho = 2$. This local minimum is a finite-$N$ effect, as we have confirmed numerically. 

\begin{figure}[t]
\includegraphics[width=9.5cm,height=4.8cm,angle=0]{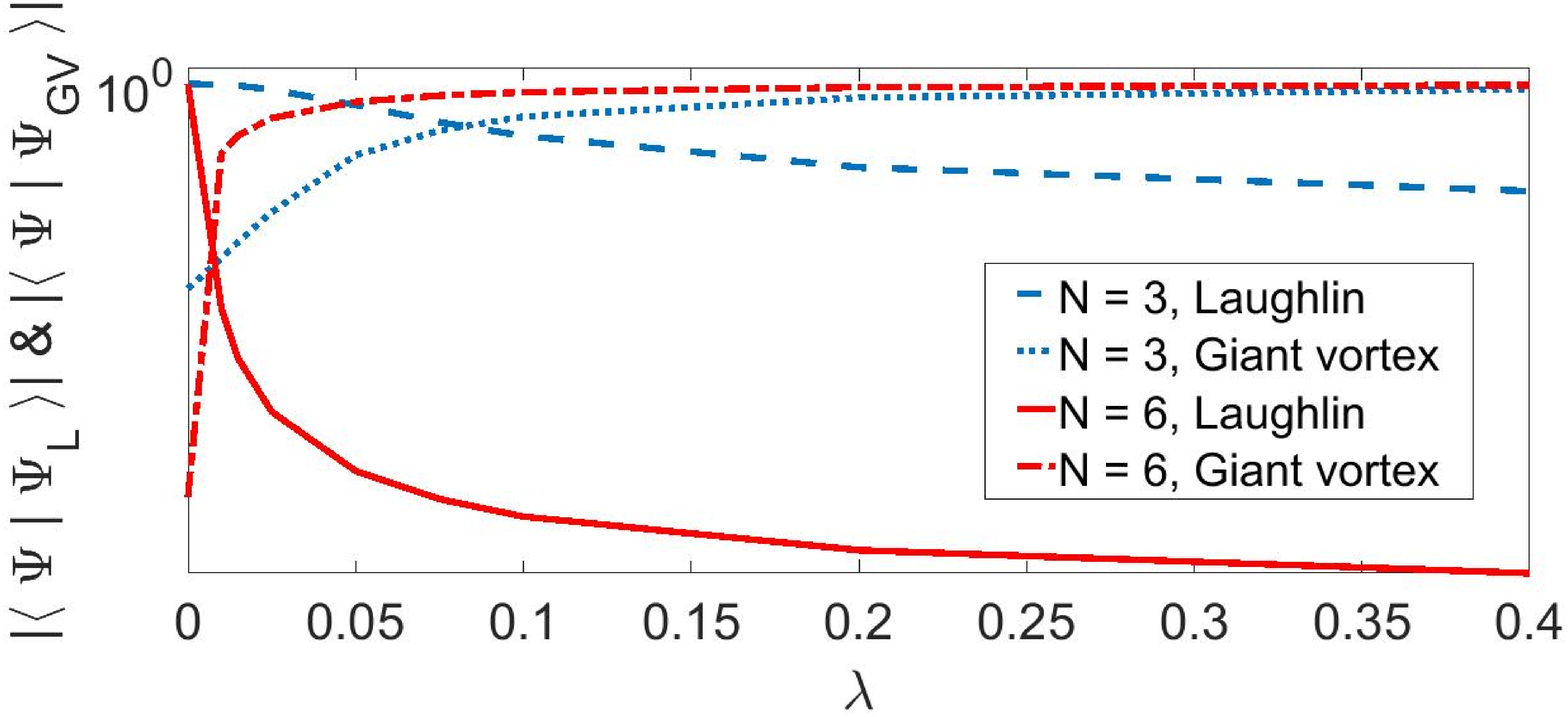}
\caption{(Color online) The overlaps $|\langle \Psi | \Psi_L \rangle|$ and $|\langle \Psi | \Psi_{\rm GV} 
\rangle|$ (on a logarithmic $y$ axis) between the lowest-energy eigenstate of the Hamiltonian and the 
Laughlin state, as well as the giant-vortex state, as a function of $\lambda$, with $(N-1) g = 0.3$. Here 
$(N,L) = (3,6)$ (dashed line for the overlap with the Laughlin state and dotted line with the giant vortex), 
and also $(N,L) = (6,30)$ (solid line with the Laughlin and dotted-dashed line with the giant vortex).}
\end{figure}

\begin{figure}[t]
\includegraphics[width=9.5cm,height=4.8cm,angle=0]{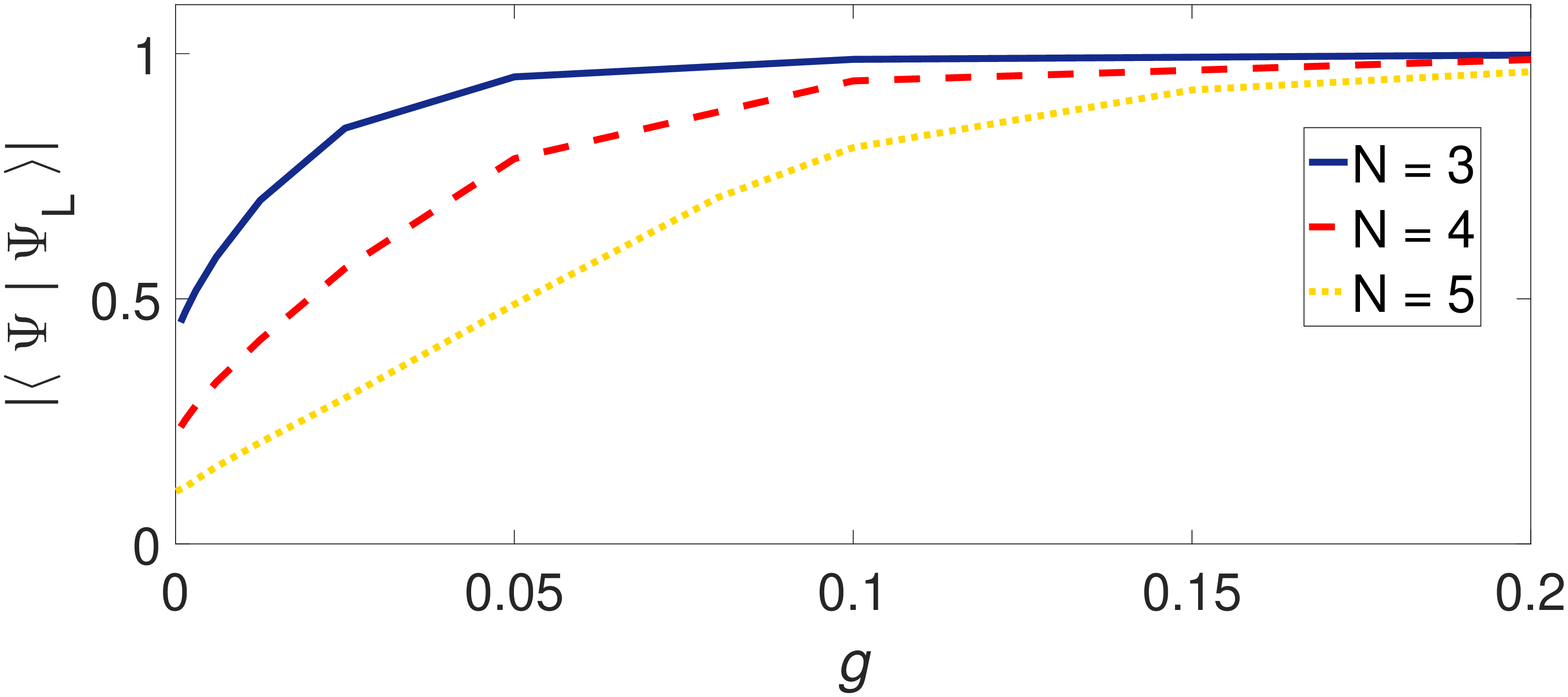}
\caption{(Color online) The overlap between the lowest-energy eigenstate of the Hamiltonian evaluated 
for $\lambda = 0.01$ and the Laughlin state (evaluated for $\lambda = 0$), as a function of $g$, for $(N,L) 
= (3,6)$ (solid line), (4,12) (dashed line), and (5,20) (dotted line).}
\end{figure}

In order to characterize the evaluated many-body state $|\Psi \rangle$, we project it on the Laughlin state 
$|\Psi_L \rangle$, as well as on the giant-vortex state $|\Psi_{\rm  GV} \rangle$. The corresponding absolute
value of the amplitudes $\langle \Psi | \Psi_L \rangle$ and $\langle \Psi | \Psi_{\rm GV} \rangle$ are shown 
in Fig.\,5 (on a logarithmic y axis) and Fig.\,6. 

In Fig.\,5 we see the effect of increasing $\lambda$, for various values of $N$, keeping the interaction 
energy of the ground state of the system $g (N-1)$ constant. It shows the transition that was mentioned 
earlier from the Laughlin state to the giant-vortex state, as $\lambda$ increases. Furthermore, $|\langle 
\Psi |\Psi_L \rangle|$ decays very rapidly with increasing $\lambda$ and $N$. For example, already for 
$N = 6$ atoms and $\lambda = 0.01$ the overlap is $\approx 0.1$. An approximate fitting formula that we 
have found for the absolute value of the slope is $N^{7.34}/e^{8.607} \approx (N/3.23)^{7.34}$. The 
extremely steep decline of the overlap implies that the Laughlin state is very fragile. More specifically, 
the anharmonicity parameter $\lambda$ has to decrease equally rapidly with increasing $N$, in order for 
the Laughlin state to survive. Turning to the overlap of $|\Psi \rangle$ with the giant vortex, $|\langle 
\Psi | \Psi_{\rm GV} \rangle|$ increases rapidly with increasing $\lambda$, and the slope also increases 
with increasing $N$, for small values of $\lambda$, as seen in Fig.\,5.

The effect of the interaction strength on the overlap of $|\Psi \rangle$ with the Laughlin state, for a 
fixed value of $\lambda$ can be seen in Fig.\,6. We observe that as the interaction strength increases, 
the effect of the anharmonic potential is suppressed and thus the overlap with the Laughlin state increases. 
Still, the question is how the overlap behaves as function of $N$. In the results shown in Fig.\,6 we evaluated 
the inner product between the Laughlin state (for $\lambda=0$) and the many-body state for $\lambda=0.01$ 
as a function of $g$, for $N=3, 4$ and 5 atoms. The interesting observation here is that the overlap 
approaches unity as $g$ increases less rapidly as $N$ increases. 

To conclude, a Bose-Einstein condensate that rotates in a purely harmonic potential undergoes a series of 
transitions as the rotational frequency of the trap increases. Singly-quantized vortex states enter the cloud, 
which eventually form a vortex lattice. When the rotational frequency approaches the trap frequency, the 
system enters a highly-correlated regime, where the number of vortices becomes comparable to the number 
of atoms. 

The question we have posed here is whether these correlated states persist in a harmonic-plus-quartic 
potential. In such a potential, within the mean-field approximation there are three distinct phases. In
the first phase we have a vortex lattice, in the second we have giant-vortex states, while the third is 
a combination of a lattice with a giant vortex which is located at the trap center. 

In the limit of rapid rotation we see that the Laughlin state competes with the giant-vortex state. The 
transition between them takes place for a ``strength" of the quartic part of the confining potential that 
decreases rapidly as the atom number increases. While we have focused on the bosonic Laughlin state [for 
$L = N (N-1)$], our results are more general and are valid, at least for the states with $L \gtrsim N(N-1)$.

The transition from the Laughlin state to the giant-vortex state may be attributed to the single-particle 
density distribution (shown in Fig.\,2) of the cloud in the two states, which is rather different in the 
two states. It is flat and extends up to a radius equal to $\sqrt{2N}$ in the Laughlin state. It has a 
Gaussian profile with a width of order unity, and its maximum is located at $\sqrt{N-1}$ in the giant-vortex 
state. The Laughlin state thus becomes energetically unfavourable, even in a weakly anharmonic trapping potential 
\cite{remark}.

The fragility of the Laughlin state makes its experimental realization virtually impossible for macroscopic atom 
numbers. For example, for a typical value of $\lambda = 10^{-3}$, as in the experiments of Refs.\,\cite{Dalibard},
$N$ should be less than roughly $N=10$ in order for the Laughlin state to be achievable. Therefore it is an
experimental challenge to realize this state, which could become possible either by reducing the atom number 
very drastically, or by decreasing the value of $\lambda$, also very drastically.


\begin{thebibliography}{55}
\bibitem{remark} In Refs.\,\cite{qhanh2} it has been argued that for $L > N (N-1)$ and under certain 
conditions, the many-body states remain correlated, with a single-particle density distribution that 
develops a hole in the middle of the trap. This is energetically favourable and is consistent with 
the results of the present study.
\end{thebibliography}
\end{document}